\documentclass{elsart3p}

\usepackage{graphicx}
\usepackage{epsfig}

\usepackage{amssymb}

\begin{document}

\begin{frontmatter}



\title{The ground state phases  of orbitally degenerate spinel oxides}


\author{George Jackeli\corauthref{cor1}}

\address{Institute for Theoretical Physics,
Ecole Polytechnique F\'ed\'erale de Lausanne, CH-1025, Lausanne,
Switzerland,\\ and E. Andronikashvili Institute of Physics, Tbilisi, Georgia.}
\corauth[cor1]{Tel: +41 21 693 44 79: fax:   +41 21 693 54 19.\\
{\it E-mail address:} george.jackeli@epfl.ch}
\begin{abstract}

I review  the microscopic spin-orbital Hamiltonian and ground state properties
of spin one-half spinel oxides with threefold $t_{2g}$ orbital degeneracy.
It is shown that for any orbital configuration  a ground state of corresponding 
spin only  Hamiltonian is infinitely degenerate in the classical limit.
The extensive classical degeneracy is lifted by the quantum nature of the spins, 
an effect similar to order-out-of-disorder phenomenon by quantum fluctuations.  This drives  the system to a non-magnetic 
spin-singlet dimer manifold with a residual degeneracy due to relative orientation of dimers. The magneto-elastic mechanism 
of lifting the ``orientational'' degeneracy is also briefly reviewed.
\end{abstract}
\end{frontmatter}
\section{Introduction}
The pyrochlore lattice which is composed by corner sharing tetrahedra, 
is known to be the most frustrated lattice existing in the nature.
For the Heisenberg antiferromagnet on  pyrochlore lattice the order-out-of-disorder mechanisms are inactive \cite{pyrtheo1} and
such a spin system would remain liquid
down to the lowest temperatures \cite{pyrtheo2}.

However, in real compounds magnetic ions, forming a frustrated lattice, often  possess an orbital degeneracy
in addition to the spin one.  In such cases, orbital degrees of freedom are also incorporated
in the superexchange theory and the systems are described by means of an effective 
Kugel-Khomskii type spin-orbital model \cite{KK}. 
The exchange interaction between magnetic moments on a bond depends now on the orientation
of occupied electronic orbitals with respect to that bond. The physical behavior of such systems is expected to
 be drastically different from that of pure spin models, as the occurrence
of an orbital ordering  can modulate the spin exchange and  partially or
fully release the geometrical degeneracy of the underlying lattice.

\begin{figure}
\centerline{\epsfig{file=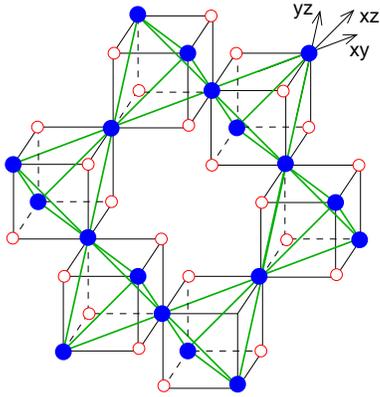,width=5.0cm}}
\caption{The pyrochlore lattice formed by B sites in the spinel structure AB$_2$O$_4$: only B (colored circles) and O 
(white circles) sites are shown.}
\label{spinel}
\end{figure}

The compounds that we have in mind  are transition metal (TM) 
spinels.  The systems which  are the subject of intense experimental 
and theoretical activity \cite{spinel}. 
All spinels have the general formula AB$_2$O$_4$ and here we will be dealing with situation
when B sites with octahedral coordination are occupied by magnetic TM ions with orbital degeneracy.
The most interesting feature of the spinel structure  is the fact that the B ions  form a  highly frustrated 
pyrochlore lattice [See Fig.\ref{spinel}].
These systems thus give the unique possibility to explore how the natural tendency of correlated systems to develop
magnetic and orbital is effected by geometrical frustration.

The degeneracy of the $d$-shell of TM ions is not fully  lifted by the ligand field with octahedral symmetry
and local electronic structure is composed by high energy $e_g$ doublet and low energy $t_{2g}$ triplet.
In the case of partial filling of the $t_{2g}$ manifold, ($d_{xy}$, $d_{xz}$, and $d_{yz}$ orbitals), 
one encounters with the  threefold orbital degeneracy  in $d^1$ and $d^2$ systems.
The first case corresponds to Ti$^{3+}$ spinels, such as MgTi$_2$O$_4$ \cite{Tiex1,Tiex2},
and the second case to V$^{3+}$ spinels  AV$_2$O$_4,$ where A=Zn, Mg, Cd \cite{Vex1,Vex2}.
Although both cases are described by similar Hamiltonians the physics of $S=1/2$ and $S=1$ systems are drastically different.
In what follows, I focus on  the  spin one-half $d^1$ system and refer the readers to Refs.\cite{V1,V2,V3} for the
discussion of $d^2$ systems. 

The theoretical studies of the effective model for titanium spinels have been performed in Refs. \cite{Ti1,Ti2}.
Here, I briefly review the obtained results, however, within a different and complimentary scheme of reasoning.

\section{Effective spin-orbital Hamiltonian}
 We assume that the low-temperature insulating phase of MgTi$_2$O$_4$ is of Mott-Hubbard type.
However, the alternative description of the insulating phase starting from the band picture 
is also possible and is given in Ref. \cite{KM}. 

In the Mott phase, an effective low energy spin-orbital Hamiltonian can be derived with the standard second-order superexchange theory
and has been reported in Refs.\cite{Ti1,Ti2}. The peculiarity of the spinel structure is due to the fact 
 that electron transfer between the nearest-neighbor (NN) sites of B-sublattice is 
governed by the direct  $dd\sigma$ overlap of $t_{2g}$ orbitals.
The $dd\sigma$ overlap  in $\alpha\beta$ plane
connects only  the corresponding orbitals of the same $\alpha\beta$ type.
The pyrochlore lattice, formed by B-ions, can also be viewed as a collection of crossing chains 
running in $xy$, $yz$, and $xz$ directions, as seen in Fig. 1.
Along the bond, for example in $xy$ direction, only the diagonal overlap between $xy$ orbitals are nonzero. 
Therefore the total number of electrons in each orbital state at a given site is
a conserved quantity and the orbital part of the effective Hamiltonian
$H$ has no dynamics. The orbital degrees are thus Potts-like $Z_3$
static variables. The spin-orbital Hamiltonian  has the following form:
\begin{eqnarray}
H&=&4J_{\rm AF}\sum_{\langle ij\rangle} {\big [}\vec S_i\cdot \vec S_{j}-\frac{1}{4} {\big ]}O^{\rm OF}_{ij}
\nonumber\\
&-&{\sum_{\langle ij\rangle}
\big [}J_{\rm O} +J_{\rm F} \vec S_i\cdot \vec S_{j}{\big ]} O^{\rm OAF}_{ij}
\label{eq1}
\end{eqnarray}
where the sum is over pairs of nearest-neighbor sites of the pyrochlore lattice.
The first term describes antiferromagnetic coupling between the NN spins and is active when the bond is occupied by the
corresponding orbitals of the same type [see Fig.2A]. For the bond $ij$ in $\alpha\beta$-plane
we have  ${O}^{\rm OF}_{ij}=P_{i,\alpha\beta}P_{j,\alpha\beta}$, where $P_{i,\alpha\beta}$ is a projector operator and is 
equal to 1 when $\alpha\beta$ orbital on site $i$ is occupied and is zero otherwise.
When the bond $ij$ in $\alpha\beta$-plane is occupied by two different orbitals, one of them being of
$\alpha\beta$ type [see Fig.2B], then there is gain of energy $J_{\rm O}$ , independent of spin configuration on the bond.
This is given by the second term of the Hamiltonian. However, in the same situation, the local Hund's coupling
$J_H$ favors the parallel orientation of the spins and introduces a weak ferromagnetic (FM) coupling $J_{F}$
 between them, described by the last term in the Hamiltonian. Along the bond in $\alpha\beta$-plane  
$O^{\rm OAF}_{ij}=P_{i,\alpha\beta}(1-P_{j,\alpha\beta})+P_{j,\alpha\beta}(1-P_{i,\alpha\beta})$. There is a local constraint
$P_{i,xy}+P_{i,xz}+P_{i,yz}=1$ as we have one electron at each site.
The coupling constants can be expressed in terms of $J=t^2/U$,  defining the  energy scale of the problem, and parameter $\eta=J_{\rm H}/U$ 
measuring the strength of the Hund's coupling with respect of local Coulomb repulsion $U$. For TM ion $\eta=J_{\rm H}/U\ll 1$ 
and is of the order of 0.1. In this limit  we have $J_{\rm AF}\simeq J_{\rm O}\sim J$, and $J_{\rm F}\sim \eta J$ \cite{Ti1,Ti2}.
\begin{figure}
\centerline{\epsfig{file=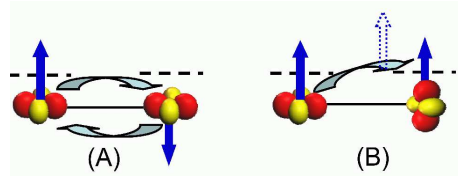,width=6.0cm,angle=0}}
\caption{Orbital arrangements on (A) antiferromagnetic and (B) ferromagnetic bonds.}
\label{bonds}
\end{figure}
\section{The ground state phases}
We start our analysis of the ground properties from well defined realistic limit $\eta=0$.
In this case the Hamiltonian Eq.(\ref{eq1}) simplifies and takes the following form:
\begin{eqnarray}
{\cal H}=E_{S}+4J\sum_{\langle ij\rangle} {\big [}\vec S_i\cdot \vec S_{j}+\frac{1}{4}{\big ]}O^{\rm OF}_{ij}
\label{eq2}
\end{eqnarray}
where $E_S=-2JN$ is the constant energy shift due to spin uncorrelated virtual fluctuations and $N$ is the number of sites.
In deriving Eq.(\ref{eq2}) we have used the fact that each lattice site has nearest-neighbors in all three directions $xy$, $xz$, and $yz$, and the constraint for projection operators given above.

The first observation is, that in the limit $\eta=0$ the only bonds occupied with corresponding same type orbitals give the contribution
to the energy. The spin exchanges on such bonds are antiferromagnetic. Different orbital configurations
will have different number and pattern of AFM exchange of underlining spin subsystems, and, in principle, different ground state energy. 
However, for classical Neel type configuration of spins on such  bonds $\langle\vec S_i\cdot \vec S_{j}\rangle=-1/4$ and
the second term in  Eq.(\ref{eq2}) vanishes. It thus follows that for any 
orbital pattern the spin subsystem have the same classical energy and the classical ground state is infinitely degenerate.

As we have noted above 
each occupied orbital has finite overlap only along corresponding chain. It is easily verified that interacting AFM bonds
can only be connected along the straight lines. Therefore, for each orbital configuration, 
the spin-subsystem can be viewed as a collection of decoupled finite (or infinite) antiferromagnetic spin one-half chains. 
The deviation  from classical
value $\langle\vec S_i\cdot \vec S_{j}\rangle=-1/4$ on AFM bond can be significant for spin one-half 
one dimensional objects. For antiferromagnetically coupled quantum spins the ground state expectation value 
 $\langle\vec S_i\cdot \vec S_{j}\rangle$ is always smaller than $-1/4$ and in this case we have finite energy gain due the second term of the Hamiltonian
Eq. (\ref{eq2}).
 The quantum energy per bond depends on the length of such spin chains. With increasing the size of AFM chain 
we have more bonds to gain the energy, however the amount of energy we gain decreases. 
The maximum energy gain per bond is for the cluster of two spins coupled into singlet state, for which we have
 $\langle\vec S_i\cdot \vec S_{j}\rangle=-3/4$. It then appears that the minimum energy configuration corresponds to such
an orbital pattern for which isolated non-interacting spin-singlet dimers are formed \cite{Ti1,jack}.
\begin{figure}
\epsfysize=136mm
\centerline{\epsffile{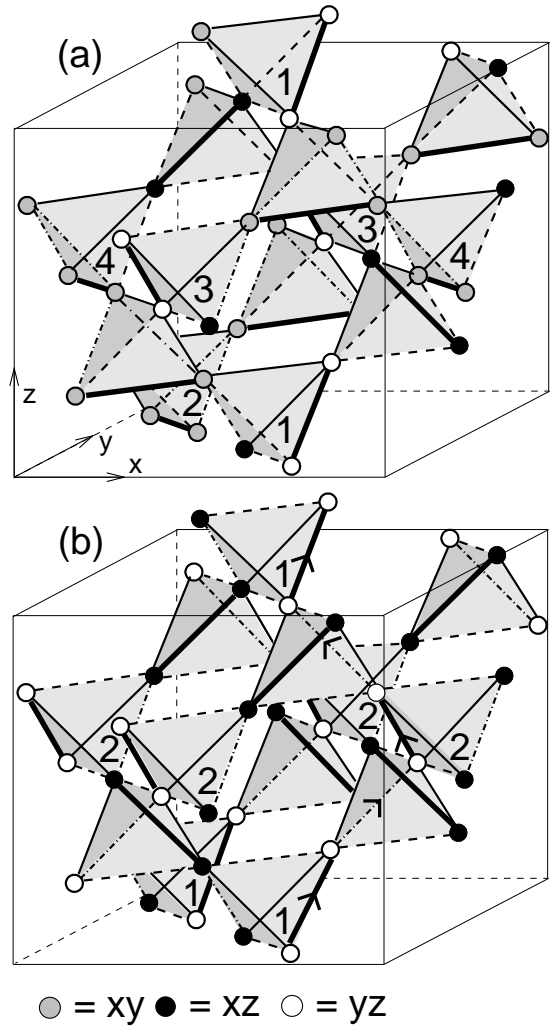}}
\caption{Two different coverings of the unit cubic cell through dimers. 
Locations of singlets are represented by thick links.
Different numbers correspond to inequivalent tetrahedra.}
\label{dimercov}
\end{figure}
The quantum nature of spins removes the spin  degeneracy and drives the system to spin-singlet nonmagnetic dimer manifold.
However, the latter is  highly degenerate with respect of dimer orientations.
One of the possible dimer coverings of the lattice is shown in Fig. 3a.
The dimers in the limit $\eta=0$ are noninteracting and the effective dimensionality of the systems is zero.
In the present case, as in a spin-Peierls system,  the  increase of magnetic energy gain due to the shortening of strong bonds
 outweights the increase in elastic energy due to the distortion of lattice. 
Therefore, each type of dimer covering induces corresponding distortion of the lattice and different distortion pattern will cost
different elastic energy. Therefore, the ``orientational'' degeneracy of dimer phase can be lifted by the elastic energy cost.
We have shown in 
Ref.\cite{Ti1,Ti2}  that  magneto-elastic interaction indeed lifts the ``orientational'' degeneracy 
and stabilizes the dimer pattern leading to the minimal enlargement of the unit cell.
 This generates a condensate of dimers in a valence bond crystal state,
forming one dimensional dimerized helical chains, indicated by arrows in Fig. 3b, running around the tetragonal $c$-axis.
Such a dimerized pattern has been  actually observed in MgTi$_2$O$_4$ \cite{Tiex2}.
There is a peculiar orbital ordering in the dimerized  phase: a ferro-type 
order along the helices  with antiferro-type order between them (see Fig. 3b).

Finally, let us briefly comment on the effect of finite Hund's coupling $\eta\not =0$. 
The latter induces the ferromagnetic coupling, $J_{\rm F}\simeq \eta J$,
between the spins belonging to different dimers. For  FM coupling much smaller than the binding energy of the spins into spin-singlet state
$J_{FM}\ll \Delta\simeq 4J$, the dimer state is stable against weak interdimer coupling. With increasing $\eta$ one finds only one phase transition, 
presumably first order, from singlet-state to a ferromagnetic state with a different orbital ordering \cite{Ti1,Ti2}.
The dimer state and ferromagnetic spin order are only possible ground state phases of spin-orbital Hamiltonian Eq.(\ref{eq1}).

\begin{ack}
I am very grateful to S. Di Matteo, C. Lacroix, and N. Perkins for  their collaborations and numerous discussions on this topic.
The support  by GNSF under the Grant No.06-81-4-100 is acknowledged.
\end{ack}


\begin{thebibliography}{9}
\bibitem{jack} The rigorous proof will be given elsewhere.
\end{thebibliography}
\end{document}